\documentclass[12pt]{amsart}
\usepackage{color}
\usepackage{algpseudocode}
\usepackage{algorithm}
\usepackage{mathtools}
\usepackage[pdftex]{graphicx}
\usepackage{epstopdf}
\usepackage{caption}
\usepackage{subcaption}
 \usepackage{amsmath}
 \usepackage{listings}
 \usepackage{hyperref}

 \usepackage{type1cm}
 \usepackage[latin1]{inputenc}
 \usepackage[T1]{fontenc}
 \usepackage{mathptmx}

\newcommand{\CPPcode}[1]{\lstinline{#1}}

\title[Multiple eigenvectors around the homo-lumo gap]{Multiple eigenvectors around the homo-lumo gap as a cheap by-product in linear scaling electronic structure calculations}
\author[A. Kruchinina]{Anastasia Kruchinina}
\date{\today}
\begin{document}
\begin{abstract}
  In this work we present and evaluate an implementation of the purify-shift-and-project method [J. Chem. Phys. 128, 176101 (2008)] for linear scaling computation of multiple eigenvectors around the homo-lumo gap of the Fock/Kohn-Sham matrix. Recursive polynomial expansions allow for linear scaling density matrix construction if matrices are sufficiently sparse. However, a drawback is that, compared to the traditional diagonalization approach, eigenvectors of the Fock/Kohn-Sham matrix are not readily available. The sharp polynomial filter, constructed in intermediate iterations of the recursive polynomial expansion, increases the relative separation of eigenvalues near the homo-lumo gap. The computed density matrix approximation is used to project away the uninteresting part of the spectrum, so that the eigenvalues of interest become the extreme eigenvalues, enabling fast convergence of a Lanczos eigensolver. We implement the purify-shift-and-project algorithm in the quantum chemistry program Ergo [SoftwareX 7, 107 (2018)]. We illustrate the performance of the method by computing 30 eigenvectors around the homo-lumo gap for large scale systems.
\end{abstract}
\maketitle


\section{Introduction}

In electronic structure calculations using Hartree-Fock~\cite{Roothaan} or Kohn-Sham density functional theory~\cite{hohen,KohnSham65}, the density matrix $D$ is computed from the Fock/Kohn-Sham matrix $F$. 
The traditional approach for computation of the density matrix involves matrix diagonalization, meaning that all eigenvectors of the Fock/Kohn-Sham matrix are explicitly computed. In recursive polynomial expansions for density matrix construction the matrix diagonalization is avoided. Provided that matrices are sufficiently sparse and large, the computational cost of recursive polynomial expansions may be reduced to linear. However, the disadvantage of this approach is that eigenvectors of the matrix $F$ are not readily available.

Eigenvectors of the Fock/Kohn-Sham matrix are used to represent molecular orbitals in a given basis set. Molecular orbitals help to understand chemical properties of molecular systems. Of particular importance are the highest occupied molecular orbital (homo) and the lowest unoccupied molecular orbital (lumo). We assume here that the homo-lumo energy gap is non-zero. 
Often, there is no need for all eigenvectors of the matrix $F$, but only a small number of interior eigenvectors around the homo-lumo gap is required.

Recently we developed and implemented in the quantum chemistry program Ergo~\cite{rudberg2018ergo} the purify-shift-and-square algorithm for computation of homo and lumo eigenvectors~\cite{kruchinina2017fly}. 
Unfortunately, the algorithm proposed in~\cite{kruchinina2017fly} can provide only the homo and lumo molecular orbitals.  In this work we implement and evaluate the performance of the purify-shift-and-project method proposed in~\cite{interior_eigenvalues_2008} for  computation of multiple eigenvectors around the homo-lumo gap of the matrix $F$. The purify-shift-and-square and purify-shift-and-project methods are inspired by the shift-and-square~\cite{morgan1991computing, wang1994electronic} and
 shift-and-project~\cite{xiang2007linear} algorithms, respectively, where the spectrum transformations are applied directly to the matrix $F$.
However, in the purify-shift-and-square and purify-shift-and-project methods, the eigenvectors are computed for matrices obtained  in intermediate iterations of the recursive density matrix polynomial expansion. The choice of recursive expansion iterations for eigenvector computations should be carefully determined with consideration of eigenvector accuracy and eigensolver performance.

The interior eigenvalue problem has been addressed in a number of works. The contour-integral based projection methods extract requested eigenpairs by solving a smaller problem in a projected subspace~\cite{imakura2016relationships,polizzi2009density,sakurai2003projection}. To construct a smaller problem, the solution of a system of linear equations with a number of right-hand sides is required. The purify-shift-and-square and  purify-shift-and-project methods are similar in spirit to other methods taking advantage of the polynomial filtering and the Lanczos process~\cite{fang2012filtered,li2016thick,schofield2012spectrum,zhou2014chebyshev}. Often, the Chebyshev approximation polynomial is used to transform the interior eigenvalue problem to an exterior one, where the eigenvalues of interest are located at the spectrum end. Usually such methods focus on the computation of a large number of eigenpairs of a given sparse Hermitian matrix. We note that our main focus is the computation of the density matrix. 
In this work the eigenvectors are provided as a by-product of the recursive polynomial density matrix expansion, which fortunately allows to construct a very sharp eigenvalue filter.

\section{Algorithm description}

Let the Fock/Kohn-Sham matrix $F$ have eigenvalues $\lambda_i$ and the  corresponding eigenvectors $y_i$:
\begin{align}
  Fy_i = \lambda_i y_i,
  \label{eq:fock}
\end{align}
where eigenvalues are arranged in increasing order 
\begin{align}
  \lambda_1 \leq \ldots \leq \lambda_{n_\text{occ}} < \lambda_{n_\text{occ}+1} \leq \ldots \leq \lambda_N,
\end{align}
and $n_\text{occ}$ is the number of occupied orbitals. Here, $\lambda_{\text{homo}} \coloneqq \lambda_{n_\text{occ}}$ is the eigenvalue corresponding to the highest occupied molecular orbital (homo), and $\lambda_{\text{lumo}}\coloneqq \lambda_{n_\text{occ}+1}$ is the eigenvalue corresponding to the lowest unoccupied molecular orbital (lumo). We assume that the homo-lumo gap $\lambda_{\text{lumo}} - \lambda_{\text{homo}} > 0$.

The density matrix $D$ is the matrix for orthogonal projection onto the subspace spanned by eigenvectors of the Fock/Kohn-Sham matrix $F$ that correspond to occupied molecular orbitals:
\begin{align}
  D = \sum_{i=1}^{n_\text{occ}} y_i y_i^T.
  \label{eq:density}
\end{align} 
The computation of the eigendecomposition~\eqref{eq:fock}, required for the density matrix construction using~\eqref{eq:density}, scales cubically with increasing system size. Recursive polynomial expansions provide an efficient way of computing the density matrix approximation by applying recursively low-order polynomials  $f_i$ to the matrix $F$:
\begin{align}
  D \approx X_n = f_n(f_{n-1}(\ldots f_1(f_0(F)) \ldots)).
\end{align}
An execution time scaling linearly with system size is obtained if the matrices are sufficiently sparse, and matrix sparsity is preserved during the iterations by removal of small matrix elements~\cite{ErrorControl}.
The initial polynomial $f_0$ usually maps the eigenvalues of $F$ into the interval $[0,\, 1]$ in reverse order.  Here we focus on the polynomials $f_i$, $i>0$, equal to the second-order spectral projection (SP2) polynomials $x^2$ or $2x-x^2$~\cite{Nikl2002}. The choice of polynomials $f_i$ in each iteration of the SP2 expansion may be based on the matrix trace~\cite{Nikl2002} or on the homo/lumo eigenvalue estimates~\cite{interior_eigenvalues_2014,kruchinina2017fly}.

Iterative eigensolvers, such as the Lanczos method, are most efficient for computation of extreme eigenpairs. Thus, usually some matrix transformation is applied to move internal eigenvalues of interest to spectrum ends. In the shift-and-square method~\cite{morgan1991computing, wang1994electronic}, the matrix spectrum is folded over itself using the  transformation $(F-\sigma I)^2$, by which the eigenvalues close to the shift $\sigma$ are moved to the end of the spectrum. Unfortunately, the folding transformation squeezes and mixes occupied and unoccupied spectrum parts, making computation of multiple eigenvectors around the homo-lumo gap very difficult.

In the shift-and-project method~\cite{xiang2007linear}, the density matrix is utilized for extracting occupied and unoccupied parts of the spectrum. The spectrum bounds $\lambda_\text{min} \leq \lambda_\text{1}$ and $\lambda_\text{max} \geq \lambda_\text{N}$ can be approximated using the Gershgorin circles theorem or the Lanczos method.
Then, occupied eigenvalues of $F$ near the homo-lumo gap correspond to the largest eigenvalues of $D(F-\lambda_\text{min}I)$, and unoccupied eigenvalues of $F$ near the homo-lumo gap correspond to the smallest eigenvalues of $(I-D)(F-\lambda_\text{max}I)$. This method preserves the separation of eigenvalues in the desired spectrum part and projects away the uninteresting spectrum part, making it attractive for computation of multiple eigenvectors around the homo-lumo gap.

The matrices $X_i$, obtained in the intermediate iterations of the recursive expansion, have the same eigenvectors as the matrix $F$. Moreover, the SP2 recursive expansion have the favorable property of separating the eigenvalues near the homo-lumo gap from each other and the rest of the spectrum, facilitating the convergence of the Lanczos eigensolver~\cite{interior_eigenvalues_2008}. 
This leads to the purify-shift-and-square and purify-shift-and-project methods originally proposed in~\cite{interior_eigenvalues_2008}.

The purify-shift-and-square method uses the transformation $(X_i-\sigma I)^2$ in a given iteration $i$ and shift $\sigma$. In~\cite{kruchinina2017fly}, recursive expansion iterations and shifts for computation of eigenvectors are chosen based on the homo-lumo estimates~\cite{interior_eigenvalues_2014}. The algorithm for selection of iterations and shifts aims to reduce the number of required Lanczos iterations. The matrix square $X_i^2$ is computed in each iteration of the recursive expansion, so it is reused in the purify-shift-and-square method. 

The purify-shift-and-project method~\cite{interior_eigenvalues_2008} uses the transformations $D(X_i-I)$ and $(I-D)X_i$ for some intermediate matrix $X_i$ for computation of occupied and unoccupied eigenvectors, respectively.  
A good density matrix approximation $D$ is available only at convergence of the recursive expansion. Therefore the computation of eigenvectors is delayed until the end of the recursive expansion, and storage of at least one intermediate matrix $X_i$ is required. We suggest to compute all requested eigenvectors in the same iteration $i$ to perform only one matrix-matrix multiplication $DX_i$. As mentioned in~\cite{kruchinina2017fly}, the matrix product $DX_i$ may be incorporated into the Lanczos procedure. However, in this case, the number of matrix-vector products will increase by a factor of two, and the eigensolver implementation need to be modified.

The separation between requested eigenvalues and the rest of the spectrum vary during the recursive expansion, but one may expect a better separation in some intermediate iteration. Finding the iteration $i$ of the recursive expansion allowing for the computation of multiple eigenvectors with good accuracy and the smallest number of Lanczos iterations is difficult. We note that the iteration $i$ should not be selected close to the convergence of the recursive expansion where numerical errors (round-off errors and errors due to the removal of small matrix elements) start to dominate the result.  In Ergo, the user defines two parameters deciding when eigenvectors will be computed. The user chooses at what iteration, counted from the end of the recursive expansion, to compute eigenvectors  using the input parameter \CPPcode{scf.go_back_X_iter_proj_method}. If in the chosen iteration the eigensolver did not converge in a requested number of Lanczos iterations, the program attempts to compute eigenvectors in some previous recursive expansion iteration. Since the purify-shift-and-project method requires storage of intermediate matrices, we provide another input parameter \CPPcode{scf.jump_over_X_iter_proj_method} deciding how many iterations to step back for a new attempt to compute eigenvectors. In this way the number of matrices saved in memory is reduced.

If only one eigenpair is required, the Lanczos eigensolver is probably the best choice since it utilizes matrix symmetry, converges fast and is rather simple to implement. However, as soon as one eigenvalue converges, Lanczos basis vectors loose orthogonality, which results in misconvergence or stagnation.
A simple fix of the orthogonality loss in the Lanczos method is full re-orthogonalization of each new Lanczos basis vector against all previous computed basis vectors. This essentially makes Lanczos method equivalent to the Arnoldi method for general matrices.

\section{Numerical experiments}

\begin{figure}[ht!]
   \centering
   \begin{minipage}[c]{.45\linewidth}
      \includegraphics[width=\textwidth]{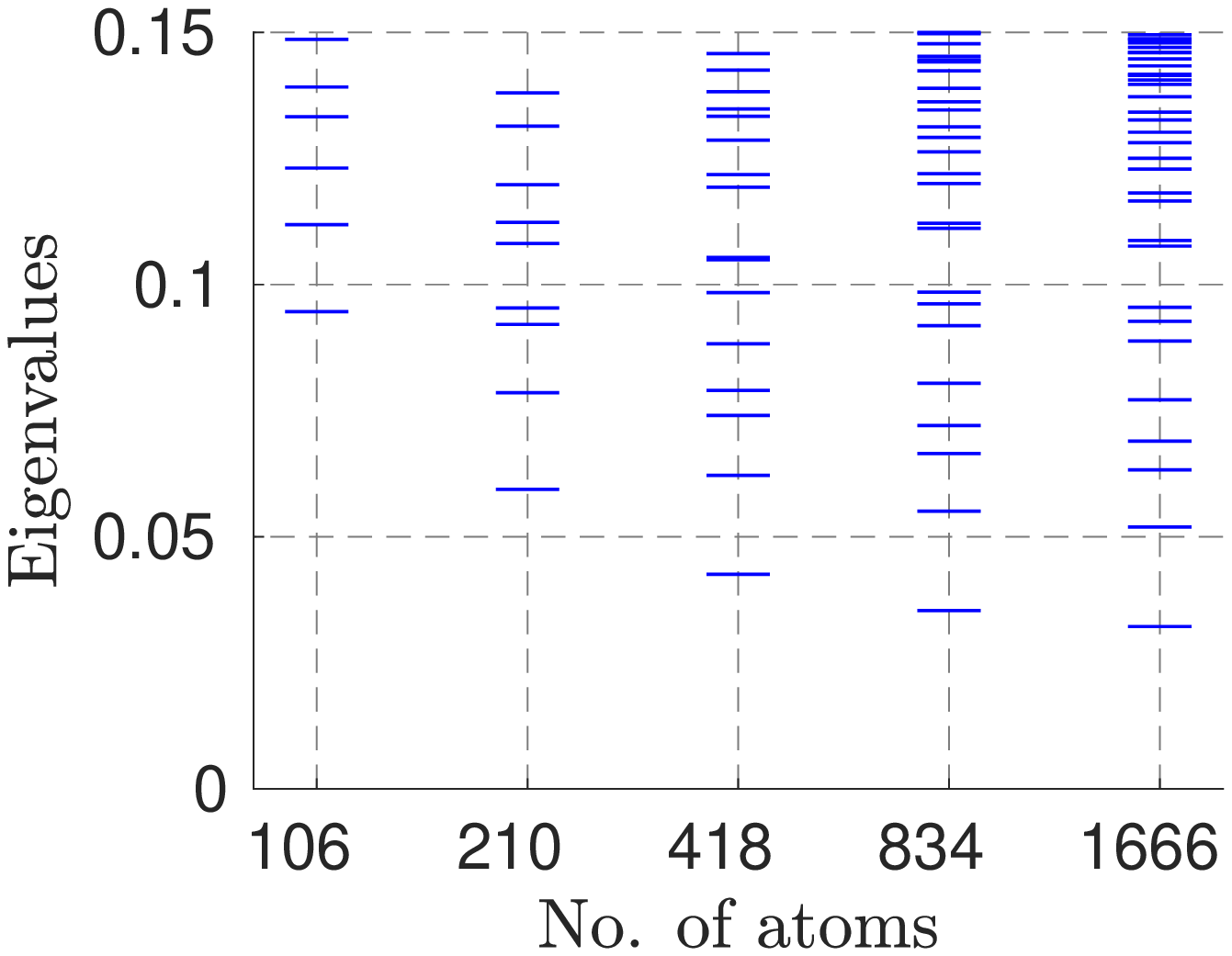}
      \subcaption{Unoccupied part of the spectrum}
   \end{minipage}
   \begin{minipage}[c]{.45\linewidth} 
      \includegraphics[width=\textwidth]{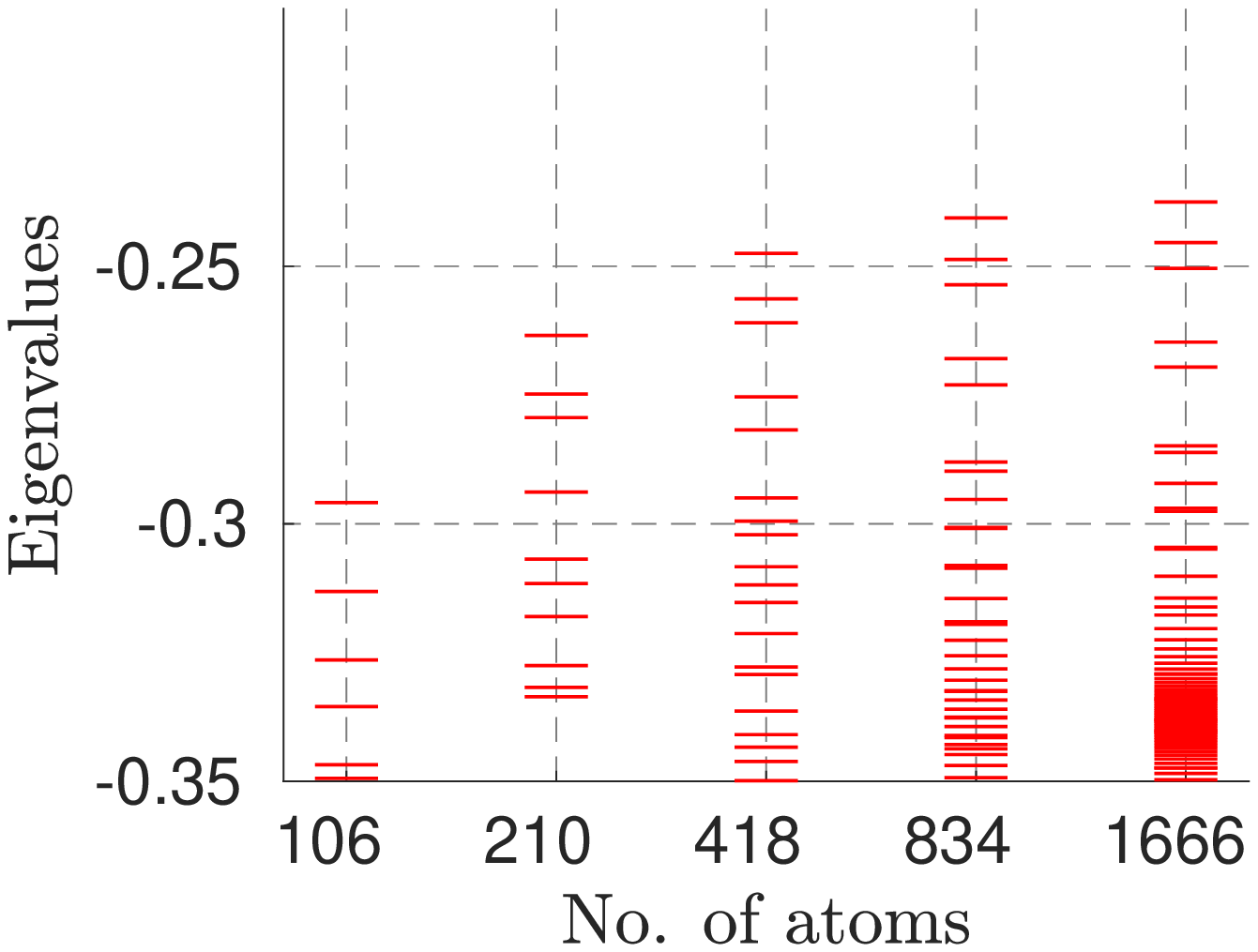}
      \subcaption{Occupied part of the spectrum}
    \end{minipage}  
    \caption{Eigenspectra near the homo-lumo gap of matrices $F$ obtained in the last SCF cycle of HF/3-21G calculations for  Glu-Ala helices of increasing size. 
    }
    \label{fig:gluala_spectrum}
  \end{figure}

Hartree-Fock calculations using the  3-21G Gaussian basis set are performed on glutamic acid and alanine (Glu-Ala) helices using quantum chemistry program Ergo~\cite{rudberg2018ergo}.
The xyz coordinates of Glu-Ala helices are available for download at \href{http://www.ergoscf.org}{www.ergoscf.org}. The largest presented system has 53250 atoms and the corresponding matrices have size 307204. Parts of the eigenspectra around the homo-lumo gap of Fock matrices obtained in the last SCF cycle for a few smaller systems are presented in Figure~\ref{fig:gluala_spectrum}. We note that the spectrum width of the Fock matrices do not change significantly with increasing system size. 
Numerical tests are performed on the Tetralith computer cluster at the National Supercomputer Centre (NSC) in Link{\"o}ping, Sweden. Each node on Tetralith is equipped with Intel Xeon Gold 6130 CPUs providing a total of 32 cores. We compiled Ergo using the Intel 18.0.3 C++ compiler and used Intel MKL version 2018.3.222 for matrix operations at the lowest level in the sparse hierarchical representation~\cite{hierarchic_2007}. We used a random initial starting guess in the Lanczos method. The Lanczos iterations were considered converged when relative residuals $\|Ax-\lambda x\|_2 / |\lambda|$ for all requested eigenpairs $(\lambda, x)$ of a given matrix $A$ were less than $10^{-12}$.

The SP2 recursive expansion is performed with control of the error in the occupied invariant subspace using the scheme presented in~\cite{ErrorControl}. We set the maximum allowed error to $7\times10^{-5}$. We use the parameterless stopping criterion developed in~\cite{stop_crit_2016} detecting when the numerical errors start to dominate and a significantly better result cannot be obtained with given settings. The polynomials are chosen based on the homo-lumo eigenvalue estimates as described in e.~g.~\cite{kruchinina2017fly}. The homo-lumo gap does not change significantly with increasing system size. Thus, the total number of recursive expansion iterations is expected to not change significantly. Indeed, for all systems 28 iterations is required to reach convergence.

\begin{figure}[ht!]
   \centering
   \includegraphics[width=\textwidth]{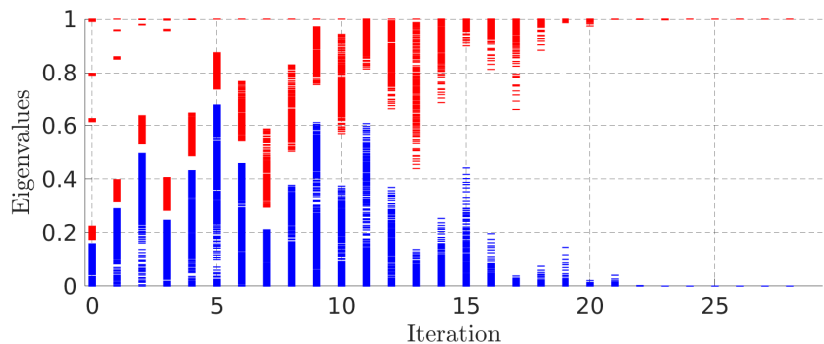}
   \caption{Eigenspectra of matrices $X_i$ in each iteration $i$ of the recursive expansion. Red points correspond to occupied eigenvalues, and blue points correspond to unoccupied eigenvalues. The matrix $F$ is obtained in the last SCF cycle of HF/3-21G calculations for the Glu-Ala helix containing 106 atoms.}
   \label{fig:spectra_X_i}
  \end{figure}

In Figure~\ref{fig:spectra_X_i} we illustrate the eigenvalue movement during the recursive expansion for the Glu-Ala helix containing 106 atoms, which corresponds to the smallest system presented in Figure~\ref{fig:gluala_spectrum}. The figure shows that the separation between eigenvalues around the homo-lumo gap is larger in intermediate iterations compared to the beginning and the end of the recursive expansion.

\begin{figure}[ht!]
   \centering
   \begin{minipage}[c]{.45\linewidth}
      \includegraphics[width=\textwidth]{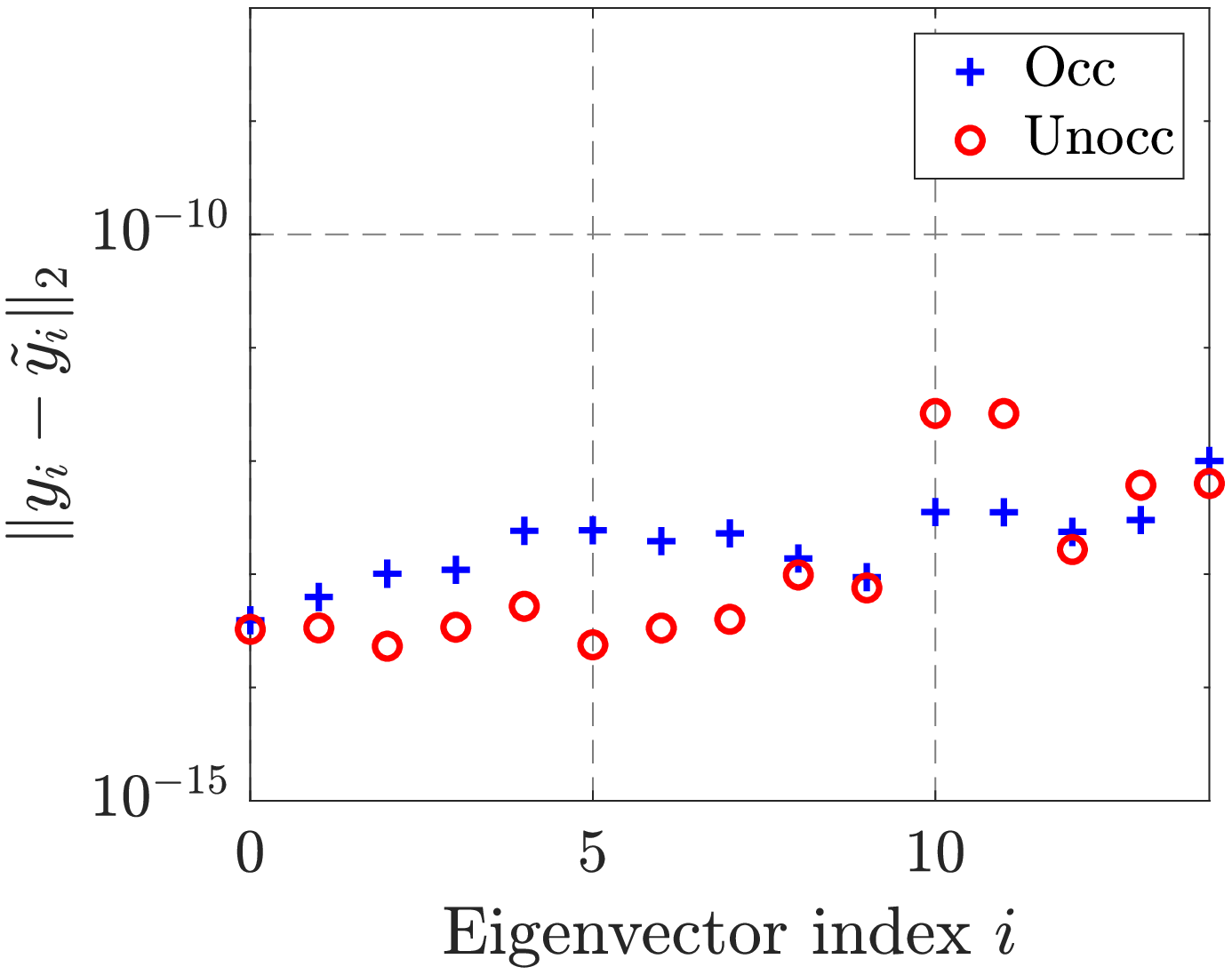}
      \subcaption{106 atoms}
   \end{minipage}
   \begin{minipage}[c]{.45\linewidth} 
      \includegraphics[width=\textwidth]{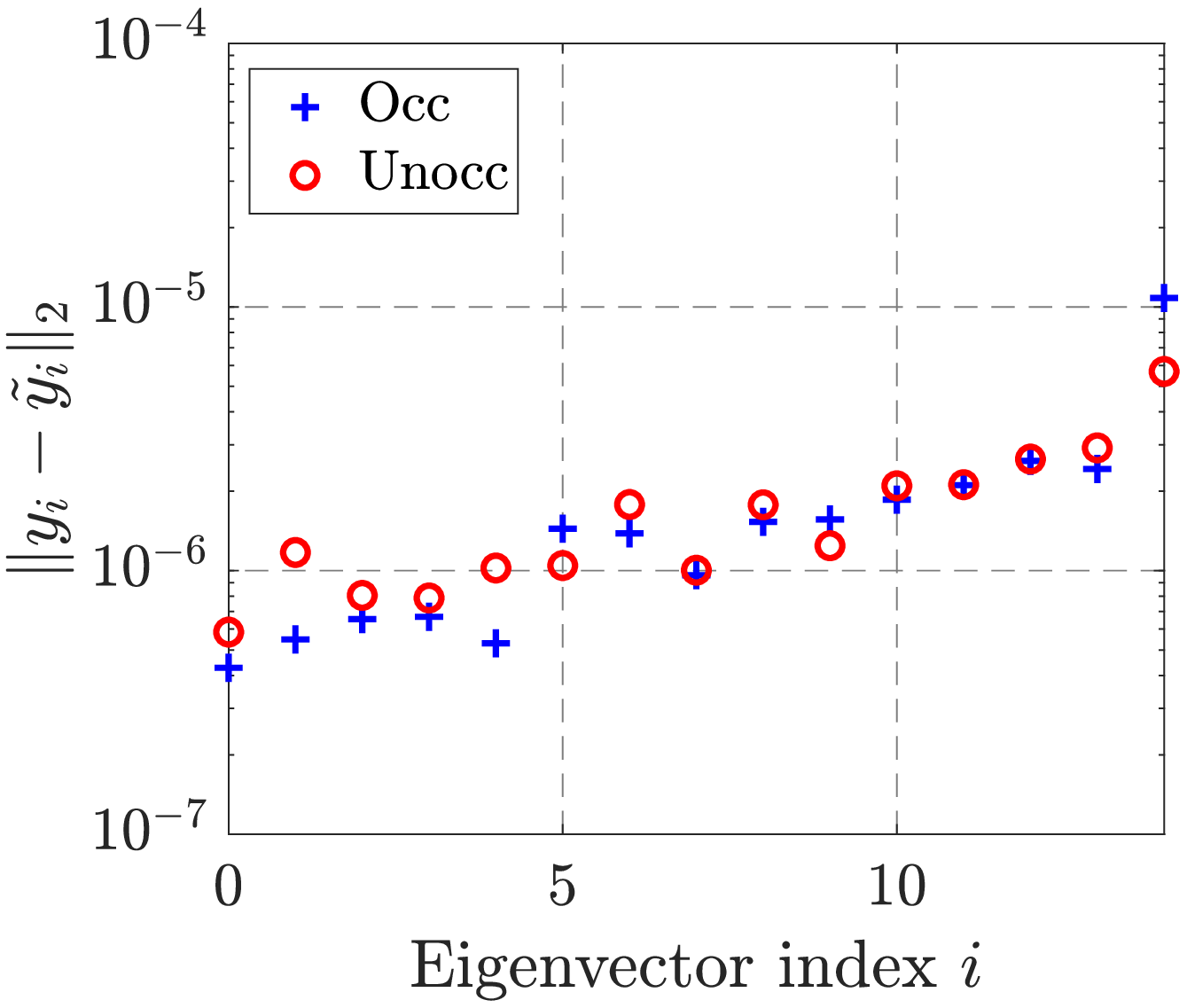}
      \subcaption{6658 atoms}
    \end{minipage}  
   \caption{Euclidean norm of the difference between eigenvectors $\tilde{y}_i$ and $y_i$ computed using the purify-shift-and-project method in iteration 18 and using full diagonalization, respectively. The index $i$ goes from 0 to 14 for both occupied (``Occ'') and unoccupied (``Unocc'') eigenvectors, where index 0 corresponds to homo and lumo eigenvectors. Matrices $F$ are obtained in the last SCF cycle of HF/3-21G calculations for the Glu-Ala helices. }
   \label{fig:error}
  \end{figure}

We use the following Ergo input parameters for the purify-shift-and-project method:\\ \CPPcode{scf.go_back_X_iter_proj_method = 10},\\ \CPPcode{scf.jump_over_X_iter_proj_method = 3}. \\Such choice of parameters leads to computation of eigenvectors using the purify-shift-and-project method in iteration 18 of the recursive expansion for all system sizes. We compute 15 occupied and 15 unoccupied eigenvectors around the homo-lumo gap using the purify-shift-and-project method. In addition, we perform a full diagonalization of the matrix $F$ using the LAPACK routine \texttt{sygv}.
In Figure~\ref{fig:error}  we  show the norm of the difference between eigenvectors computed in the  purify-shift-and-project method and eigenvectors obtained by the full
 diagonalization for systems containing 106 and 6658 atoms. The corresponding matrix sizes are 604 and 38404. The matrix $F$ obtained for the smaller system is dense.

\begin{figure}[ht!]
     \centering
      \includegraphics[width=0.5\textwidth]{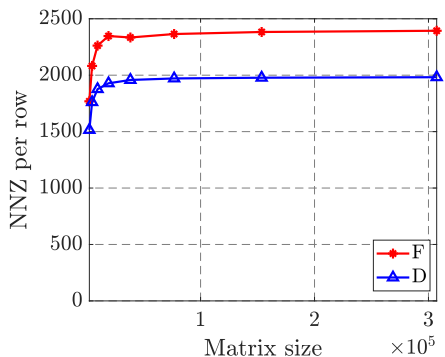}
     \caption{The number of non-zero elements per row in Fock (F) and density matrices (D). Matrices $F$ of increasing size are obtained in the last SCF cycle of HF/3-21G calculations for Glu-Ala helix systems. The smallest presented system has 418 atoms and the corresponding matrix size is 2404. The largest presented system has 53250 atoms and the corresponding matrix size is 307204.}
     \label{fig:nnz}
    \end{figure}

\begin{figure}[ht!]
 \centering
   \includegraphics[width=0.7\textwidth]{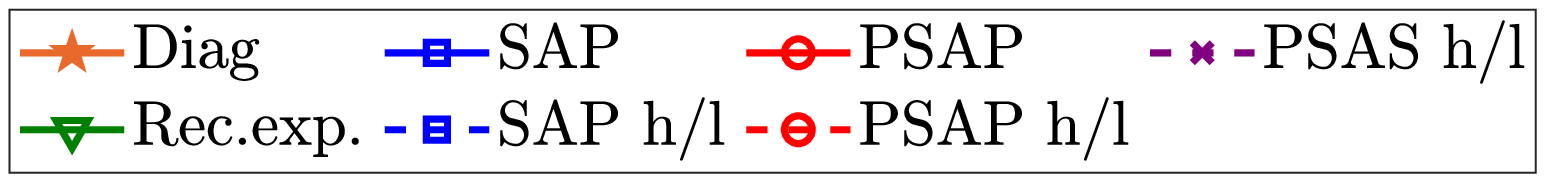}
 
   \includegraphics[width=0.49\textwidth]{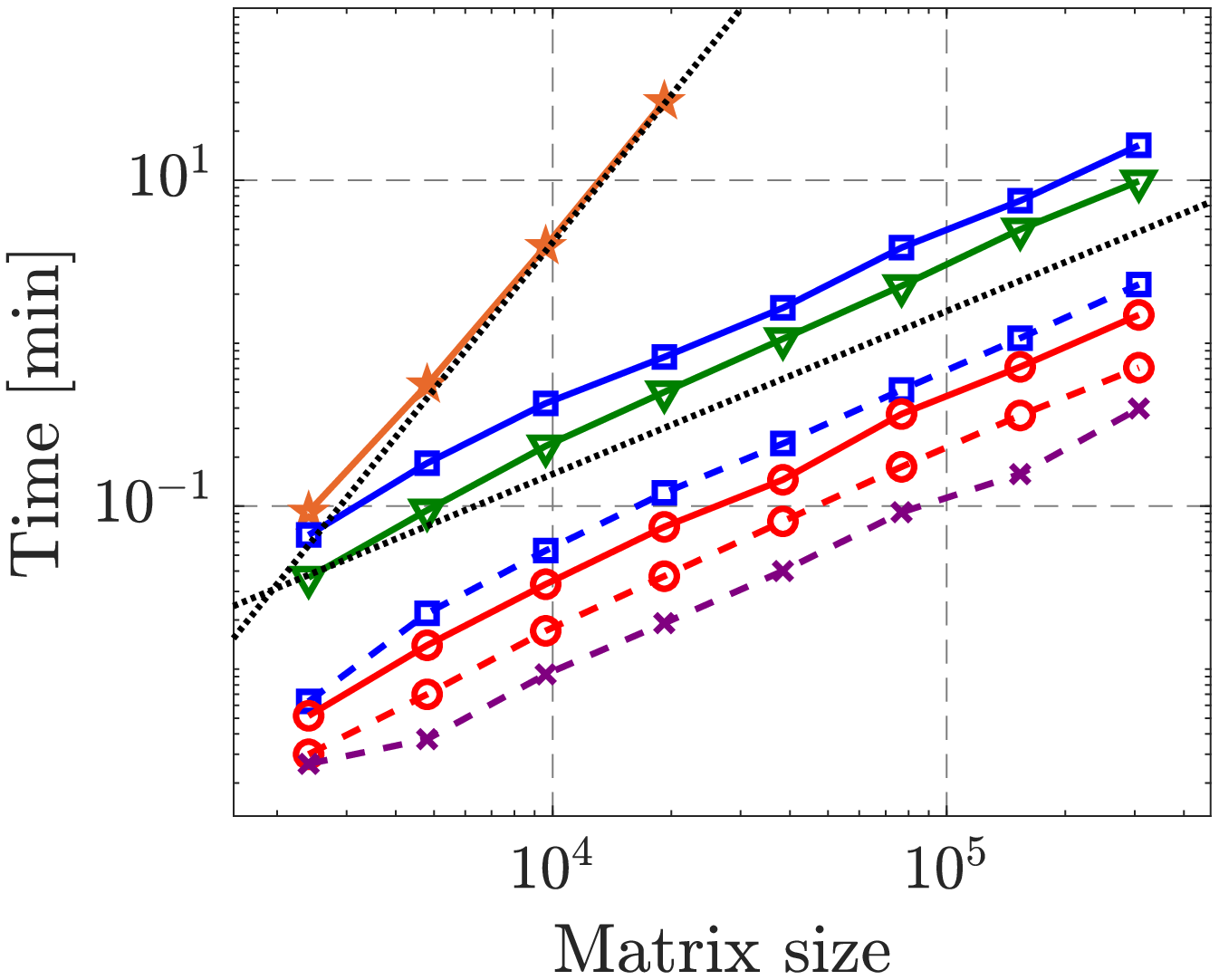}
   \includegraphics[width=0.49\textwidth]{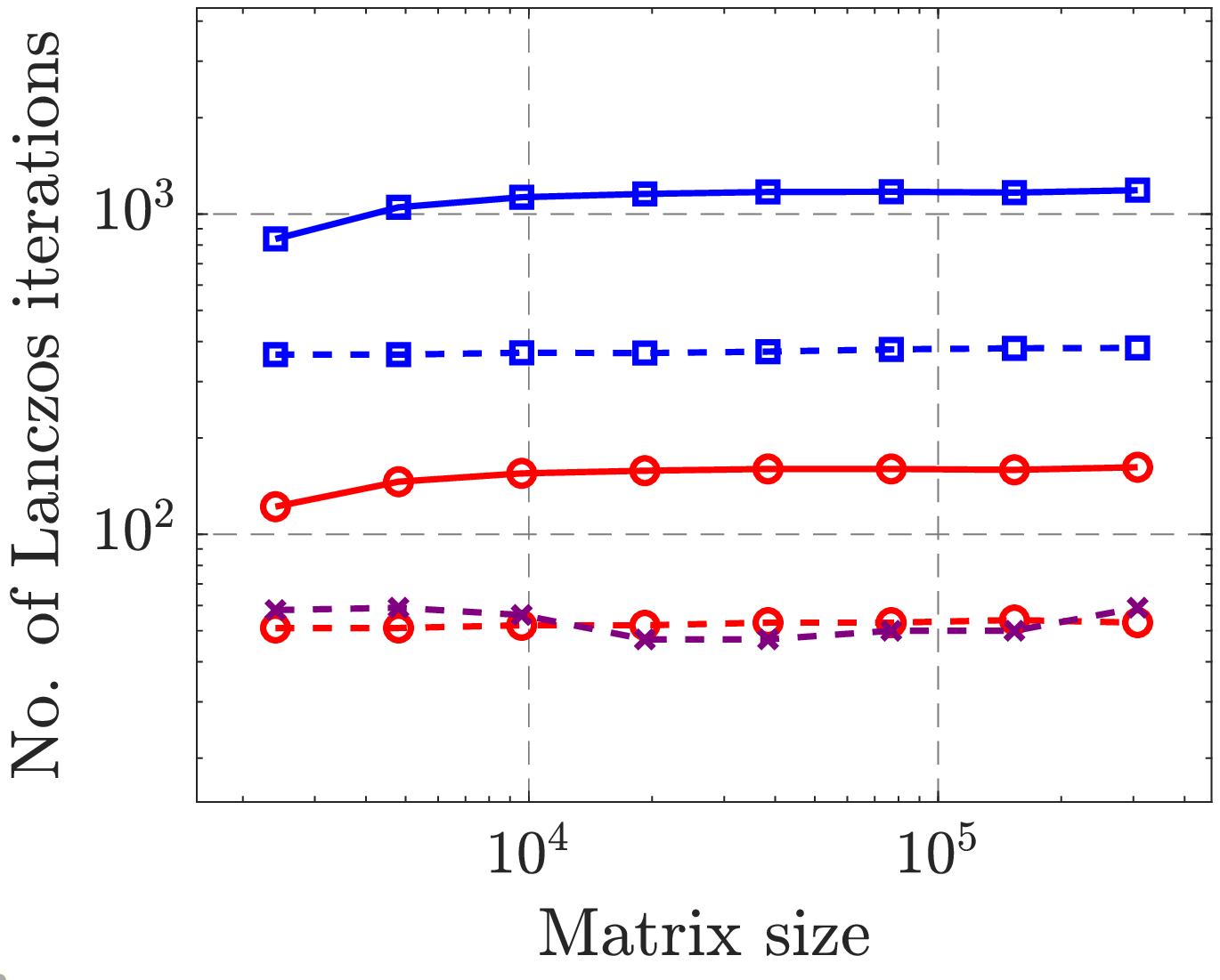}
 \caption{\underline{Left panel}: execution times of the full diagonalization of the matrix $F$ using LAPACK routine \texttt{sygv} ( ``Diag''), the  SP2 recursive expansion excluding computation of eigenvectors (``Rec.exp.''), computation of 15 occupied and 15 unoccupied eigenvectors using the purify-shift-and-project method (``PSAP''), computation of 15 occupied and 15 unoccupied eigenvectors using the shift-and-project method (``SAP''), computation of homo and lumo eigenvectors using the purify-shift-and-project method (``PSAP h/l''), computation of homo and lumo eigenvectors using the shift-and-project method (``SAP h/l'') and computation of homo and lumo eigenvectors using the purify-shift-and-square method (``PSAS h/l''). 
 Dotted lines show linear and cubic trend lines. \underline{Right panel}: number of Lanczos iterations required for computation of all requested eigenvectors. Note the log-log scale in both panels.
 }
 \label{fig:timing_all}
\end{figure}

The number of non-zero elements per row in the Fock matrices $F$ and obtained density matrices $D$ remains constant for sufficiently large system sizes, as shown in Figure~\ref{fig:nnz}. 
In Figure~\ref{fig:timing_all} we compare the execution times for the full diagonalization of the matrix $F$ using the LAPACK routine \texttt{sygv}, total time spent for the SP2 recursive expansion without computation of eigenvectors, and time spent on computation of eigenvectors by different methods. In addition, we present the total number of Lanczos iterations required for computation of all requested eigenvectors, both occupied and unoccupied. The figure shows that the number of iterations for each particular method does not change significantly with increasing system size. 

We compare the performance of the purify-shift-and-project and shift-and-project methods for computation of 15 occupied and 15 unoccupied eigenvectors of $F$. We apply the shift-and-project method to the matrix $X_{0}$ obtained before the start of the recursive expansion. We note that the convergence of the Lanczos method is affected by the relative eigenvalue separation~\cite{GolubMatrixComp}, which does not change by the transformation of the matrix $F$ to the initial matrix $X_0$. 
We also compare both methods for computing only homo and lumo eigenvectors of $F$. Independently on the number of requested eigenvectors, the purify-shift-and-project method shows smaller execution time than the shift-and-project method. This is due to the favorable eigenvalue distribution obtained in later recursive expansion iterations facilitating the Lanczos eigensolver convergence.

In the purify-shift-and-square method presented in~\cite{kruchinina2017fly}, the shift and the recursive expansion iterations used for homo and lumo eigenvector computations are carefully selected based on homo and lumo eigenvalue estimates.  For the system presented in Figure~\ref{fig:spectra_X_i} chosen iterations are 17 for homo computation and 19 for lumo computation.  The number of Lanczos iterations in the purify-shift-and-square and purify-shift-and-project methods for computation of homo and lumo eigenvectors is almost the same, oscillating around 50-60 iterations for all system sizes. However, due to the matrix-matrix product $DX_i$, the execution time of the purify-shift-and-project method is larger than the execution time of purify-shift-and-square. We note that the number of Lanczos iterations in the purify-shift-and-project method may be reduced by computing eigenvectors in later iterations of the recursive expansion. However, the matrix-matrix multiplication $DX_i$ will give a significant contribution to the execution time independently on the chosen iteration $i$. 
We do not show any results for the shift-and-square method, since regardless of what shift in the homo-lumo gap is used, the Lanczos eigensolver did not
converge within 1000 iterations.

The computational cost of the density matrix construction in Ergo scales linearly for sufficiently sparse and large matrices~\cite{kruchinina2017fly,rudberg2018ergo}. Unfortunately, there is no guarantee that the computation of multiple eigenvectors will scale linearly too. The convergence of the Lanczos eigensolver is affected by the eigenvalue distribution, and convergence is fast if eigenvalues of interest are well separated from each other and the rest of the spectrum. If distinct eigenvalues are very close to each other, a large number of eigensolver iterations is expected. However, if the number of required Lanczos iterations does not change significantly with the system size, linear scaling computation of multiple eigenvectors around the homo-lumo gap is possible.

\section{Conclusions}

\begin{sloppypar}
A key feature of the purify-shift-and-project method~\cite{interior_eigenvalues_2008} is usage of the recursive expansion as an eigenvalue filter, providing a better relative separation between eigenvalues near  the homo-lumo gap. The computed density matrix is used to convert the original internal eigenvalue problem into an external one, which is favorable for the Lanczos eigensolver. We compared performance of the shift-and-project and purify-shift-and-project methods for computation of  multiple eigenvectors around the homo-lumo gap. In addition, we compared the performance of the shift-and-project, shift-and-square, purify-shift-and-project and purify-shift-and-square methods for computation of only homo and lumo eigenvectors. We have shown that the purify-shift-and-project method allows efficient computation of many non-degenerate homo and lumo molecular orbitals, providing a good alternative to full diagonalization of the Fock/Kohn-Sham matrix. If only the homo and lumo eigenvectors are required, the purify-shift-and-square method is superior to other methods.
\end{sloppypar}

\section*{Acknowledgments}

The author  would like to thank Assoc. Prof. Emanuel H. Rubensson for comments and suggestions that greatly improved the manuscript.

Computational resources were provided by the Swedish National
Infrastructure for Computing (SNIC) at the National Supercomputer Centre (NSC)
in Link{\"o}ping, Sweden.

\bibliography{biblio} \bibliographystyle{siam} 

\end{document}